\documentclass[prl,singlepage,showpacs,superscriptaddress,groupedaddress]{revtex4}  % for double-spaced preprint
\usepackage{graphicx}
\usepackage{epstopdf}
\usepackage{amsmath,amsthm,amssymb}
\usepackage{color}
\usepackage{amsfonts}
\usepackage{subfigure}
\usepackage{bm}
\usepackage{epstopdf}

\begin{document}

% The following information is for internal review, please remove them for submission
\widetext
%\leftline{Version 1 as of \today}
%\leftline{Primary authors: }
%\leftline{To be submitted to PRL }
%\leftline{Comment to {\tt d0-run2eb-nnn@fnal.gov} by xxx, yyy}
%\centerline{\em D\O\ INTERNAL DOCUMENT -- NOT FOR PUBLIC DISTRIBUTION}

% the following line is for submission, including submission to the arXiv!!
%\hspace{5.2in} \mbox{Fermilab-Pub-04/xxx-E}

\title{Experimental observation of controllable kinetic constraints in a cold atomic gas}
%\input author_list.tex       % D0 authors (remove the first 3 lines
                             % of this file prior to submission, they
                             % contain a time stamp for the authorlist)
                             % (includes institutions and visitors)

                                                         % contain a time stamp for the authorlist)

\author{M.M. Valado}
\affiliation{Dipartimento di Fisica ``E. Fermi'', Universit\`a di Pisa, Largo Bruno Pontecorvo 3, 56127 Pisa, Italy}
\affiliation{INO-CNR, Via G. Moruzzi 1, 56124 Pisa, Italy}

\author{C. Simonelli}
\affiliation{Dipartimento di Fisica ``E. Fermi'', Universit\`a di Pisa, Largo Bruno Pontecorvo 3, 56127 Pisa, Italy}
\affiliation{INO-CNR, Via G. Moruzzi 1, 56124 Pisa, Italy}

\author{M. D. Hoogerland}
\affiliation{Department of Physics, University of Auckland, Private Bag 92019, Auckland, New Zealand}

\author{I.Lesanovsky}
\affiliation{School of Physics and Astronomy, University of Nottingham, Nottingham, NG7 2RD, United Kingdom}

\author{J.P. Garrahan}
\affiliation{School of Physics and Astronomy, University of Nottingham, Nottingham, NG7 2RD, United Kingdom}

\author{E. Arimondo}
\affiliation{Dipartimento di Fisica ``E. Fermi'', Universit\`a di Pisa, Largo Bruno Pontecorvo 3, 56127 Pisa, Italy}
\affiliation{INO-CNR, Via G. Moruzzi 1, 56124 Pisa, Italy}
\affiliation{CNISM UdR Dipartimento di Fisica ``E. Fermi'', Universit\`a di Pisa, Largo Pontecorvo 3, 56127 Pisa, Italy}

\author{D. Ciampini}
\affiliation{Dipartimento di Fisica ``E. Fermi'', Universit\`a di Pisa, Largo Bruno Pontecorvo 3, 56127 Pisa, Italy}
\affiliation{INO-CNR, Via G. Moruzzi 1, 56124 Pisa, Italy}
\affiliation{CNISM UdR Dipartimento di Fisica ``E. Fermi'', Universit\`a di Pisa, Largo Pontecorvo 3, 56127 Pisa, Italy}

\author{O. Morsch}
\affiliation{Dipartimento di Fisica ``E. Fermi'', Universit\`a di Pisa, Largo Bruno Pontecorvo 3, 56127 Pisa, Italy}
\affiliation{INO-CNR, Via G. Moruzzi 1, 56124 Pisa, Italy}

\date{\today}

\begin{abstract}
Many-body systems relaxing to equilibrium can exhibit complex dynamics even if their steady state is trivial. At low temperatures or high densities their evolution is often dominated by steric hindrances affecting particle motion [1,2,3]. Local rearrangements are highly constrained, giving rise to collective - and often slow - relaxation. This dynamics can be difficult to analyse from first principles, but the essential physical ingredients are captured by idealized lattice models with so-called kinetic constraints [4]. Here we experimentally realize a many-body system exhibiting manifest kinetic constraints and measure its dynamical properties. In the cold Rydberg gas used in our experiments, the nature of the constraints can be tailored through the detuning of the excitation lasers from resonance [5,6,7,8], which controls whether the system undergoes correlated or anti-correlated dynamics.  Our results confirm recent theoretical predictions [5,6], and highlight the analogy between the dynamics of interacting Rydberg gases and that of soft-matter systems. 

\end{abstract}

\maketitle
Complex collective relaxation in many-body systems is often accompanied by a dramatic slowdown of diffusion processes and the emergence of non-ergodic and glassy phases [1,2,3]. These features can be seen to be the consequence of effective kinetic constraints in the dynamics [9]. A kinetic constraint is a condition on the rate for a local transition dependent on the local environment: the transition and its reverse - irrespective of whether they are energetically favourable or unfavourable - can only occur if the constraint is satisfied. Kinetic constraints can severely restrict relaxation in situations where local particle arrangements make satisfying them unlikely, which is typical of fluid systems with excluded volume interactions such as dense colloids or supercooled liquids.  When a constraint is satisfied, however, the transition is allowed and a local rearrangement is "facilitated" [4,9].  Kinetic constraints naturally give rise to collective and spatially heterogeneous relaxation, and are used to describe situations where the correlation properties of the dynamics go beyond those of the static stationary state, a salient feature of glassy systems [3].  While kinetic constrains and dynamic facilitation are argued to play a central role in the behaviour of glass formers [9], it is difficult unambiguously to establish a connection between microscopic processes and emerging kinetic constraints. \\

In our experimental study we report the emergence of manifest kinetic constraints in the many-body dynamics of laser-driven gases of Rydberg atoms in the presence of noise. We explore two kinds of constraints which lead to the suppression or the facilitation of atomic excitations in the vicinity of atoms excited to Rydberg states. Our experimental technique allows us to measure both the mean number of excitations and the fluctuations around the mean, and we find clear signatures of the constraints in the pronounced (anti-) correlations evident in both quantities.  The experimental observations are in excellent agreement with numerical simulations of a kinetically constrained system of Ising spins, where the constraint is explicitly derived from the actual interatomic interactions.\\

The kinetic constraints and resulting correlated excitation processes explored in this paper are shown schematically in Fig. 1. We consider a many-body system consisting of atoms that can be represented as pseudo Ising spins having a (non-interacting) ground state $|g\rangle$ (spin down) and an excited state $|r\rangle$ (spin up) coupled by a Rabi frequency $\Omega$. Atoms at positions $r_i$ and $r_j$ in the Rydberg state $|r\rangle$  interact through the van der Waals interaction $V_{ij}= \frac{C_6}{|r_i-r_j|^6}$,  where $C_6$ is the van der Waals interaction coefficient [10,11] (which, in the present experiment, is positive). Here we explore the regime of incoherent evolution, $\gamma>>\Omega$,  where $\gamma$ is the decay rate of the atomic coherences. In this strong dissipation limit the evolution of the many-body system is restricted to the subspace of classical spin configurations by virtue of the quantum Zeno effect [12]. The dynamics can be written as a classical rate equation with (de-)excitation rates $\Gamma_i(\Delta)= \frac{\Omega^2}{2 \gamma}\frac{1}{1+\left(\frac{\Delta-\frac{1}{\hbar}\sum_{i\neq j}V_{ij}nj}{\gamma}\right)^2}$ [5,7], where $n_j=1$ if the atom at position $r_j$ is in the Rydberg state and $0$ otherwise. Generally, therefore, the excitation rate $\Gamma_i(\Delta)$ for an atom depends on the state of all the atoms in its vicinity and on the detuning $\Delta$ of the excitation laser from resonance. For $\Delta = 0$ (in practice, $\Delta < \gamma $), the interactions between an atom  and its neighbours lead to a blockade constraint  resulting in anti-correlated dynamics: the more excited atoms there are in the vicinity, the smaller $\Gamma_i$, leading to a spatially inhomogeneous local excitation rate and an overall slowing down of the dynamics as the number of excitations in the systems grows.  The inter-particle distance below which this \emph{blockade constraint} becomes important is given by the (incoherent) blockade radius $r_b=(C_6/\hbar \gamma)^{\frac{1}{6}}$ .  By contrast, for $\Delta > 0$ the single-atom excitation rate is small, but now a \emph{facilitation constraint} appears: an excited atom somewhere in the system shifts atoms contained within a shell of radius $r_{fac}$  and width $\delta r_{fac}$ into resonance, where  $r_{fac}=(C_6/\hbar \Delta)^{\frac{1}{6}}$, which corresponds to the van der Waals interaction compensating the laser detuning [8,13,14,15,16,17], and $\delta r_{fac}= \frac{r_{fac}}{6 \Delta}\gamma$. This facilitation leads to strongly correlated dynamics.\\

In our experiments the above model is realized with $^{87}$Rb atoms in a magneto-optical trap (MOT) that are excited to repulsively interacting 70S Rydberg states [18] and detected by a channeltron after field ionization (see Methods). In order to verify that the dynamics of our system is in the regime of validity of the above rate equation, we first measure the resonant excitation dynamics, represented by the mean number of excitations $N$ as a function of time $t$ (here and in the rest of the paper we report the derived actual quantities obtained by dividing the observed quantities by the detection efficiency), for different values of the Rabi frequency $\Omega$. In the incoherent regime $(\gamma >> \Omega)$ a scaling of $\Gamma_i$, and hence of the overall dynamics, with $ \Omega^2 / \gamma$ is expected (as opposed to the coherent regime, where the theoretical linear scaling with $\Omega$ was demonstrated in [19]). Fig. 2 shows the results of experiments for three different values of $\Omega$. When plotting the experimental data  versus $t\cdot\Omega^2  / \gamma$ all three curves collapse onto a single curve, demonstrating the expected scaling of the incoherent dynamics. \\

We now proceed to investigate the constrained non-equilibrium dynamics of our system. To realize the blockade constraint we set $\Delta=0$ and measure $N$ as a function of time. In order not to saturate our detection system for long excitation times (which, in practice, means keeping the observed number of excitations $< 40$) we use the 1D configuration (see Methods). The degree of correlation of our system is controlled by the number of atoms per blockade length $\frac{r_b}{a}$ (where $a=\left(\frac{V_{exc}}{N_g}\right)^{\frac{1}{3}}$ is the mean distance between $N_g$ ground state atoms in the excitation volume $V_{exc}$, and $r_b= 11.1\,\mathrm{\mu m}$  for our parameters), which we vary by changing the effective density of the MOT (see Methods). In this way, we can prepare samples with atom numbers corresponding to values of $\frac{r_b}{a}$ between around 1.3 (i.e.,  close to the non-interacting case  $\frac{r_b}{a}\leqslant 1$ ) and  $\frac{r_b}{a}=4.2$. The results of these experiments  are shown in Fig. 3 a), together with a numerical simulation (see Methods) that exhibits excellent qualitative and good quantitative agreement (to within overall factors between 0.5 and 2, which are indicated in the captions to Figs. 3 and 4). The cross-over between the initial excitation regime in which the interactions play no role, reflected by a linear increase in $N$, and the blockade regime is clearly visible, indicating the point at which the average distance $d=\left( \frac{V_{exc}}{N}\right)^{\frac{1}{3}}$ between excitations becomes smaller than $r_b$ [20]. When the results are normalized by the number of ground state atoms $N_g$ in the excitation volume, which yields the fraction of excited atoms,  in the blockade regime  the excited fraction levels off much more sharply for large values of $\frac{r_b}{a}$. \\

The above observations can be represented more quantitatively by examining the average growth rate of excitations per atom, $(d N/dt)/N_g$ , as a function of $d$. The growth rate is extracted by numerically differentiating the $N$ vs $t$ data from Fig. 3 a) after smoothing them in order to avoid artefacts due to noisy data. We expect this rate to decrease sharply with decreasing values of d below the blockade radius, which defines the effective range of influence of a Rydberg atom. This is clearly borne out by experiment, as can be seen in Fig. 3 c), in which all the data sets follow a single curve which is flat for $d>r_b$ and decreases by four orders of magnitude between $11\,\mathrm{\mu m}$ and $6\,\mathrm{\mu m}$. We compare this universal curve to the theoretically predicted average single-atom growth rate derived from the rate function $\Gamma_i (\Delta=0)$ in a mean-field approach [5] and find good agreement.\\

The strong correlations caused by the kinetic constraints are expected to also be clearly visible in the fluctuations of the systems, which we investigate through the Mandel $Q$-parameter [21]. Similarly to the dipole blockade in the coherent regime [22,23], the effect of the blockade constraint is expected to reflect itself in a negative $Q$-parameter that decreases with the number of excitations since for large numbers the system has fewer choices for distributing the excitations, which in turn leads to reduced fluctuations. Moreover, following the above reasoning, the $Q$-parameter should depend exclusively on the number of excitations and not on the number of ground state atoms, so plotting $Q$ as a function of $N$ for all three data sets should yield a single curve. Again, this is experimentally confirmed in Fig. 3 d).\\

In contrast to the blockade constraint, which causes strong anti-correlations in the dynamics, the facilitation constraint should lead to a strongly correlated evolution. In order to explore this regime, we now set $\Delta/2\pi =+19\, \mathrm{MHz}$, for which $r_{fac} = 6.4\, \mathrm{\mu m}$ and $\delta r_{fac} = 39\, \mathrm{n m}$. Since we expect the predicted facilitation dynamics to be the more pronounced the larger the overall facilitation volume, which grows with an increasing number of excitations, we choose the 3D configuration for this experiment. The results are shown in Fig. 4 (a).  In this figure, three stages can clearly be distinguished: the initial nucleation stage (for $ t< 10\, \mathrm{\mu s}$) in which $N$  grows slowly due to off-resonant single particle excitations that are suppressed by a factor $\frac{1}{1+(\Delta/\gamma)^2} \approx1.4 \times 10^{-3}$ compared to the resonant regime, and in most cases no seed excitation has yet been created to nucleate the facilitation process; the facilitation stage ($10\, \mathrm{\mu s} <t< 50\, \mathrm{\mu s}$) in which the number of excitations grows fast due to successive facilitation starting from an initial seed; and a saturation stage ($t>50\, \mathrm{\mu s}$), in which the dynamics decelerates again due to the finite size of the atomic cloud (this regime is visible in the experimental data but already affected  by spontaneous decay, which is not included in the simulations; the lifetime of the 70S state is around $150\, \mathrm{\mu s}$). These regimes are also reflected in a plot of the average total growth rate as a function of time (Fig. 4(b)), in which the acceleration of the dynamics at around $10\, \mathrm{\mu s}$ can be clearly seen.   \\

To demonstrate that the sign of the detuning is crucially important, we repeated the above experiment for $\Delta/2\pi =-19 \, \mathrm{MHz}$ (Fig. 4(a); for comparison with the blockaded constraint, data for $\Delta =0$ are also shown). Here, the facilitation condition cannot be fulfilled, and only off-resonant excitation of single Rydberg atoms is possible. \\

Apart from the acceleration of the dynamics in the facilitation stage, the strong correlations induced by the facilitation constraint are clearly discernible in the Mandel $Q$-parameter shown in Fig. 4 (c). For $\Delta/2\pi =+19 \, \mathrm{MHz}$, $Q$ grows up to $30\, \mathrm{\mu s}$, becoming large and positive, and then decreases to around 0. This can be understood as follows: in the facilitation regime an excited atom enables the probabilistic excitation of  further atoms, so that small fluctuations in the timing of the excitation are amplified for short times, leading to a large value of $Q$. For long times, the dynamics becomes similar to the blockaded case as no further facilitation is possible, and hence $Q$ decreases. This non-trivial behaviour of the correlations is a genuine many-body effect and, therefore, absent for negative detuning. The fact that $Q$ is not 0 in that case, as it should be for a purely Poissonian process, is probably due to technical noise and fluctuations in the atom number, laser intensity and detuning.\\

We have shown that manifest kinetic constraints govern the dynamics of Rydberg gases.  The constraints are a consequence of quantum mechanics, but in the strong dissipation regime studied here the dynamics is effectively classical.  Future experiments could systematically probe the crossover from the incoherent to the coherent regime, giving insight into the role of quantum effects in constrained dynamic relaxation. Furthermore, techniques for spatially resolving Rydberg excitations [24,25,26] should reveal rich spatial correlations in the dynamics [27], in analogy with dynamic heterogeneity in glasses [3].  More generally, our results give a first indication of the broad potential for applying ideas from soft-matter physics to the study of interacting atomic systems.\\

\section{Acknowledgments}
The research leading to these results has received funding from the European Research Council under the European Union's Seventh Framework Programme (FP/2007-2013) through the Marie Curie ITN COHERENCE and the ERC Grant Agreement n. 335266 (ESCQUMA), as well as through the H2020 FET Proactive project RySQ (grant N. 640378) and the EPSRC Grant no. EP/J009776/1.

\section{Methods}

\subsection{Experimental protocol}

We create cold samples of $^{87}$Rb atoms in a magneto-optical trap. The resulting roughly spherical atomic clouds have a Gaussian density profile with widths between $45$ and $160 \, \mathrm{\mu m}$ and contain between $10^4$ and $1.2\times10^6$ atoms. We control the (effective) number of atoms and hence the density of the clouds (keeping their size and shape constant) by depumping a fraction of the atoms in the $F$=2 hyperfine ground state to the $F$=1 state using a resonant laser pulse of duration up to $2 \, \mathrm{\mu s}$ (for details see [28]).  This state lies $6.8\, \mathrm{GHz}$ below the $F$=2 state (the ground state for the Rydberg excitation process) and hence does not couple to the Rydberg excitation lasers. After the depumping pulse, we infer the effective atom number by exciting a few Rydberg atoms (less than $10$, in order to avoid interaction effects) on resonance and  measuring the total growth rate, which in the non-interacting, incoherent regime is proportional to the atom number.\\

Excitation to the 70$S$ Rydberg state (for which $C_6 = h \times 1337 \, \mathrm{GHz \, \mu m^6}$ ) is achieved by a two-photon process with laser beams near $420\, \mathrm{nm}$ (waist $10$ or $40 \, \mathrm{\mu m}$) and $1013\, \mathrm{nm}$ (waist $110\, \mathrm{\mu m}$), and the $420\, \mathrm{nm}$ laser detuned by $660\, \mathrm{MHz}$ from the $6P_{3/2}$ excited state. Depending on the sizes and propagation directions of those beams, we can realize two geometries: and effective 1D geometry, in which the $420\, \mathrm{nm}$ laser is focused to a waist of around $10 \, \mathrm{\mu m}$ (comparable to the blockade and facilitation radii) and the $1013\, \mathrm{nm}$ intersects the $420\, \mathrm{nm}$ beam and the atomic cloud at a $45$ degree angle; and a 3D geometry, in which the $420\, \mathrm{nm}$ laser has a waist of $40 \, \mathrm{\mu m}$ and co-propagates with the $1013\, \mathrm{nm}$ laser. The combined decoherence rate due to the laser linewidths and residual Doppler broadening is estimated to be $\gamma= 2\pi \times 0.7\, \mathrm{MHz}$.\\

After an excitation pulse of up to  $100\, \mathrm{\mu s}$, the Rydberg atoms are field ionized and detected by a channeltron with an overall detection efficiency of around $40\%$. The experiments are repeated $100$ times for each set of parameters in order to obtain the mean number and variance.

\subsection{Numerical simulation }

The dynamics of a laser driven Rydberg gas in the limit of strong dissipation is given by the equation [12] $\dot{\nu}=\sum_i \Gamma_i(|r_i\rangle \langle g_i| +|g_i\rangle \langle r_i| -|r_i\rangle \langle r_i| -|g_i\rangle \langle g_i|) \nu$ where $\nu$ encodes the probability of the system to be in one of its classical configurations (products of the local spin states $|r_i\rangle$ and $|g_i\rangle$). In order to account for the inhomogeneity of the atomic cloud we distributed the atomic positions according to a Gaussian profile in close approximation to the experimental situation. The spatially inhomogeneous laser profile was accounted for by using a local varying Rabi frequency for the calculation of the value of the rate function $\Gamma_i$ for each individual atom. In the simulation only atoms inside an effective excitation volume $V_{exc}$ were considered. This excitation volume was defined as the region in which the free flipping rate of an atom was at least $10\%$ (3D data) or $5\%$ (1D data) of the maximum rate. The simulations where conducted by implementing a classical kinetic Monte Carlo algorithm, and the results were averaged over $1000$ runs (1D data) or $100$ runs (3D data).
\\
\subsection{Mean-field calculation of the excitation rate }

The theoretical curve in Fig. 3 d) was obtained from the expression for $\Gamma_i(\Delta$)  in a mean-field approach by averaging $ \Omega^2 $ over the excitation volume defined above and substituting $|r_i-r_j|$ by half the mean distance between excitations, where the volume used for calculating the mean distance was chosen as $1.92 \times 10^4  \, \mathrm{\mu m^3}$, which gave the best agreement with experiment. The interaction term in $\Gamma_i(\Delta$)   was multiplied by $6$ in order to account for six nearest neighbours for each atom.

\clearpage
\section{Figures}

\begin{center}
\includegraphics[width=13cm]{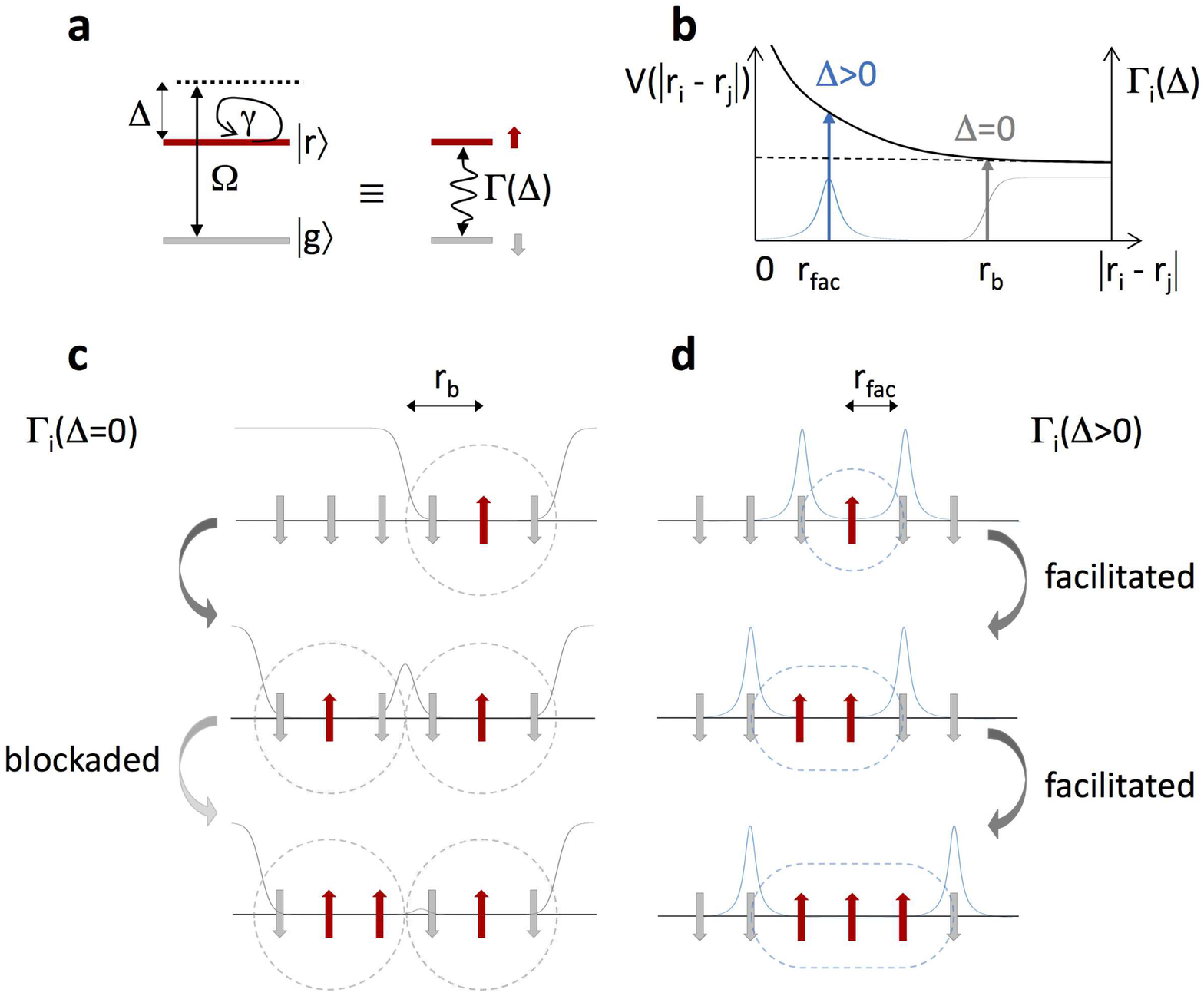}
\end{center}
\textbf{Figure 1} Kinetic constraints realised in a gas of Rydberg atoms. a) Schematic representation of the laser-induced coupling between ground and Rydberg states with Rabi frequency $\Omega$ and detuning $\Delta$, where coherence between the atomic states is lost at a rate $\gamma$. Each atom can be effectively described as an Ising pseudo spin. In the incoherent regime $\gamma>> \Omega$, the dynamics of an atom reduces to incoherent state changes (spin flips) at a rate $\Gamma_i (\Delta)$. b) Interaction potential $V(|r_i-r_j|)$ between two excited atoms at positions $r_i$ and $r_j$ (black line) and excitation rates $\Gamma_i (\Delta)$ for $\Delta>0$ (blue line) and $\Delta=0$ (grey line). The excitation rate for $\Delta=0$ drops to zero for interatomic distances $|r_i -r_j |$ below the blockade radius $r_b$ (\emph{blockade constraint}), whereas it peaks at the facilitation radius $r_{fac}$  for $\Delta>0$ (\emph{facilitation constraint}). This leads to blockaded dynamics, shown in c), and to facilitated dynamics, shown in d), respectively. In c), excitation of individual atoms occur at the resonant rate until the distance between adjacent excitations approaches the blockade radius. The third spin from the left is excited at a strongly reduced rate as it is within the blockade radius. In d), a seed excitation enables further excitations at the facilitation distance $r_{fac}$.\\

\begin{center}
\includegraphics[width=13cm]{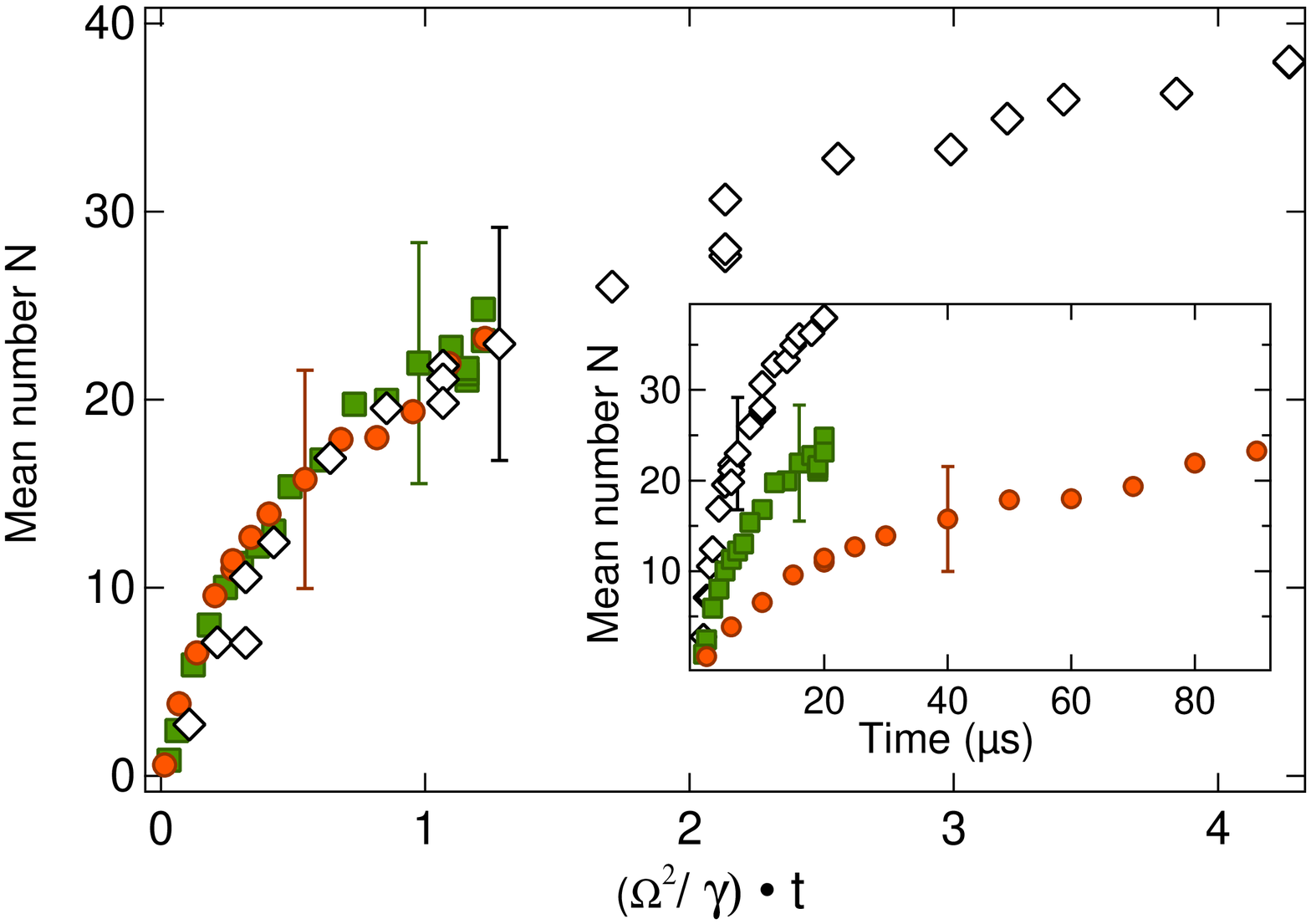}
\end{center}
\textbf{Figure 2} Dependence of incoherent dynamics on the Rabi frequency. The inset shows the mean number of excitations in a sample of $7\times10^5$ atoms in a MOT of width $60\, \mathrm{\mu m}$ in the 3D configuration as a function of time for three different Rabi frequencies: $\Omega /2\pi = 81$ (open diamonds), $43$ (green squares) and $20$ kHz (red circles). When multiplying the excitation times by $ \Omega^2 / \gamma$ (where $\gamma/2\pi = 0.7\, \mathrm{MHz}$), the three curves collapse onto each other (main figure), demonstrating the expected $\Omega^2$ scaling in the incoherent excitation regime. Representative error bars are one s.d.m.\\

\begin{center}
\includegraphics[width=13cm]{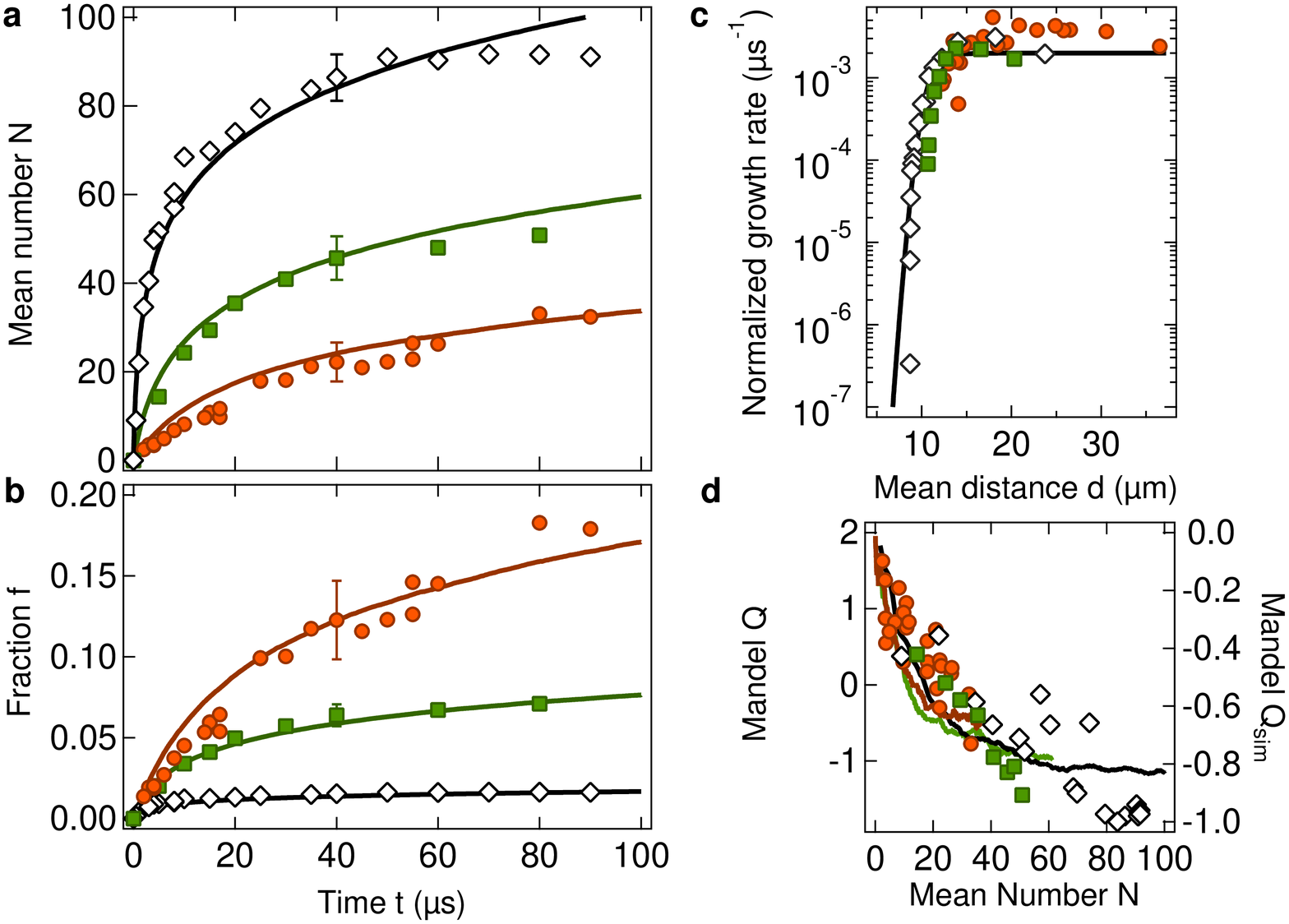}

\end{center}
\textbf{Figure 3} Evidence for the blockade constraint in a gas of Rydberg atoms. a) Mean number of excitations and b) excitation fraction as a function of time for different atom numbers $N_g$  inside the interaction volume: $5600$ (open diamonds), $715$ (green squares) and $180$ (red circles). These correspond to values of the parameter $\frac{r_b}{a}$ of $4.2$, $2.1$ and $1.3$, respectively. The numerical simulations (solid lines) have been  scaled vertically by a factor  of $1.8$. c) Average single-particle excitation rate as a function of the mean distance $d$ between excited atoms. The dashed line is obtained from the expression for $\Gamma_i(\Delta)$  in a mean-field approach (see Methods). d) Mandel $Q$-parameter as a function of mean excitation number. The vertical axis on the left corresponds to experimental data, that on the right to the numerical simulation (solid lines). The deviations of the experimental data from the simulation are most likely due to experimental noise (positive $Q$ for small values of $N$) and saturation effects of the detection system (values of $Q<-1$ for large $N$). Representative error bars are one s.d.m.\\

\begin{center}
\includegraphics[width=13cm]{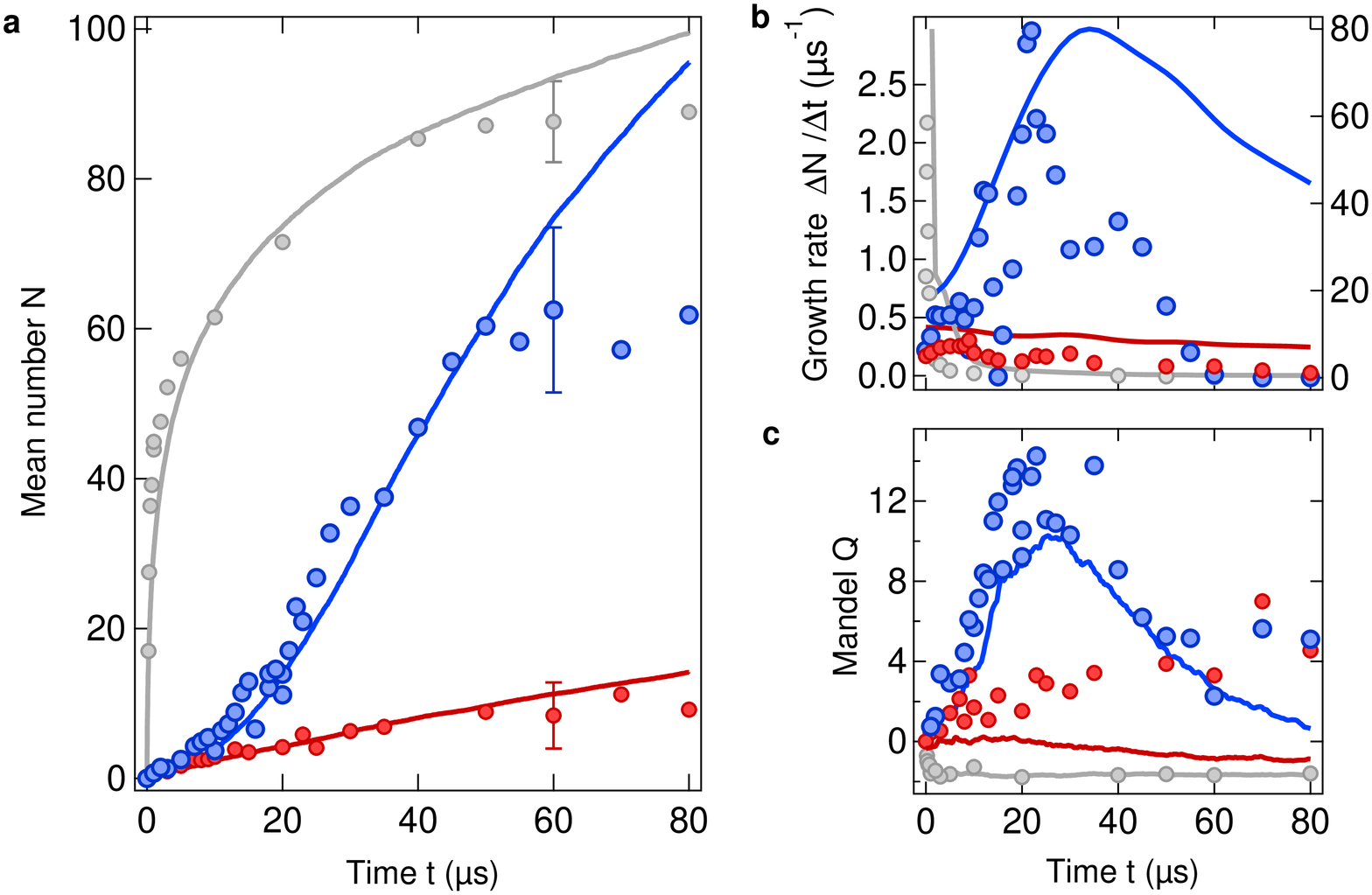}
\end{center}
\textbf{Figure 4} Evidence for the facilitation constraint in a gas of Rydberg atoms. a) Mean number of excitations as a function of time for $\Delta/2\pi =+19\, \mathrm{MHz}$ (blue circles), $\Delta/2\pi =-19\, \mathrm{MHz}$ (red circles) and $\Delta=0$ (grey circles). The numerical simulations (solid lines) have been scaled vertically by a factor $0.56$. Representative error bars are one s.d.m. b) The growth rate of excitations as a function of time. The left-hand vertical axis refers to the data for $\Delta/2\pi =+19\, \mathrm{MHz}$ and $\Delta/2\pi =-19\, \mathrm{MHz}$, the left-hand axis refers to $\Delta=0$ . The numerical simulations (solid lines) have been scaled vertically by a factor $0.56$.  c) The Mandel $Q$-parameter as a function of time. The numerical simulations (solid lines) have been scaled by a factor $2$.\\

%\bibliographystyle{apsrev4-1}
%\bibliographystyle{unsrtnat} % Use the "unsrtnat" BibTeX style for formatting the Bibliography

%\bibliography{BibliographyVdW_V11} % The references (bibliography) information are stored in the file named "Bibliography.bib"

\clearpage
 \section{References}
 
[1] Binder, K. $\And$ Kob, W. Glassy Materials and Disordered Solids:  An Introduction to Their Statistical Mechanics (World Scientific, Singapore, 2011). \\

[2] Ediger, M.D. $\And$  Harrowell, P. Perspective: Supercooled liquids and glasses. J. Chem. Phys. \textbf{137}, 080901 (2012).\\

[3] Biroli, G. $\And$  Garrahan, J.P. Perspective: The Glass Transition, J. Chem. Phys.  \textbf{138}, 12A301 (2013).\\

[4] Ritort, F. $\And$  and Sollich, P. Glassy dynamics of kinetically constrained models, Adv. Phys.  \textbf{52}, 219 (2003).\\

[5] Lesanovsky, I. $\And$   Garrahan, J.P. Kinetic Constraints, Hierarchical Relaxation, and Onset of Glassiness in Strongly Interacting and Dissipative Rydberg Gases. Phys. Rev. Lett.  \textbf{111}, 215305 (2013).\\

 [6] Lesanovsky, I. $\And$   Garrahan, J.P. Out-of-equilibrium structures in strongly interacting Rydberg gases with dissipation. Phys. Rev. A  \textbf{90}, 011603(R) (2014).\\

[7] Hoening, M., Abdussalam, W., Fleischhauer, M. $\And$   Pohl, T. Antiferromagnetic long-range order in dissipative Rydberg lattices. Phys. Rev. A  \textbf{90}, 021603(R) (2013).\\

[8] G\"{a}rttner, M., Heeg, K.P., Gasenzer, T. $\And$  Evers, J. Dynamic formation of Rydberg aggregates at off-resonant excitation. Phys. Rev. A  \textbf{88}, 043410 (2013).\\

[9] Chandler, D. $\And$  Garrahan, J.P. Dynamics on the Way to Forming Glass: Bubbles in Space-time. Annu. Rev. Phys. Chem.  \textbf{61}, 191 (2010). \\

[10] Comparat, D. $\And$  Pillet, P. Dipole blockade in a cold Rydberg atomic sample. J. Opt. Soc. Am. B  \textbf{27}, A208 (2010).\\

[11] L\"{o}w, R. \emph{et al.} An experimental and theoretical guide to strongly interacting Rydberg gases. J. Phys. B: At. Mol. Opt. Phys.  \textbf{45}, 113001 (2012).\\

[12] Marcuzzi, M., Schick, J., Olmos, B. $\And$  Lesanovsky, I. Effective dynamics of strongly dissipative Rydberg gases. J. Phys. A: Math. Theor.  \textbf{47}, 482001 (2014).\\

[13] Ates, C., Pohl, T., Pattard, T. $\And$  Rost, J. M.  Many-body theory of excitation dynamics in an ultracold Rydberg gas. Phys. Rev. A  \textbf{76}, 013413 (2007).\\

[14] Carr, C., Ritter, R., Wade, C. G., Adams, C. S. $\And$  Weatherill, K. J. Nonequilibrium phase transition in a dilute Rydberg ensemble. Phys. Rev. Lett.  \textbf{111}, 113901 (2013).\\

[15] Schempp, H.  \emph{et al.} Full Counting Statistics of Laser Excited Rydberg Aggregates in a One-Dimensional Geometry. Phys. Rev. Lett.  \textbf{112}, 013002 (2014).\\

[16] Malossi, N.  \emph{et al.} Full Counting Statistics and Phase Diagram of a Dissipative Rydberg Gas. Phys. Rev. Lett.  \textbf{113}, 023006 (2014).\\

[17] Urvoy, A.  \emph{et al.} Strongly Correlated Growth of Rydberg Aggregates in a Vapor Cell. Phys. Rev. Lett.  \textbf{114}, 203002 (2015).\\

[18] Gallagher, T. F. Rydberg Atoms (Cambridge University Press, Cambridge, 1994).\\

[19] Heidemann, R.  \emph{et al.} Evidence for Coherent Collective Rydberg Excitation in the Strong Blockade Regime. Phys. Rev. Lett.  \textbf{99}, (2007).\\

[20] Guti\'errez, R.,  Garrahan, J.P.,  $\And$   Lesanovsky, I. Self-similar non-equilibrium dynamics of a many-body system with power-law interactions.  arXiv:1507.02652 (2015).\\

[21] Viteau, M.  \emph{et al.} Cooperative Excitation and Many-Body Interactions in a Cold Rydberg Gas. Phys. Rev. Lett.  \textbf{109}, 053002 (2012).\\

[22] Urban, E.  \emph{et al.} Observation of Rydberg blockade between two atoms. Nature Physics  \textbf{5}, 110 (2009).\\

[23] Ga\"{e}tan,  \emph{et al.} Observation of collective excitation of two individual atoms in the Rydberg blockade regime. Nature Physics  \textbf{5}, 115 (2009).\\

[24] Schwarzkopf, A., Sapiro, R. E., $\And$  Raithel, G. Imaging Spatial Correlations of Rydberg Excitations in Cold Atom Clouds. Phys. Rev. Lett.  \textbf{107}, 103001 (2011).\\

[25] Schau{\ss}, P.  \emph{et al.} Observation of spatially ordered structures in a two-dimensional Rydberg gas. Nature (London)  \textbf{491}, 87 (2012).\\

[26] G\"{u}nter, G.  \emph{et al.} Observing the Dynamics of Dipole-Mediated Energy Transport by Interaction-Enhanced Imaging. Science  \textbf{342}, 954 (2013).\\

[27] Petrosyan, D., H\"{o}ning, M. $\And$  Fleischhauer, M. Spatial correlations of Rydberg excitations in optically driven atomic ensembles. Phys. Rev. A  \textbf{87}, 053414 (2013).\\

[28] Valado, M.M.  \emph{et al.} Preparing cold atomic samples in different density regimes for Rydberg studies. J. Phys.: Conf. Ser.  \textbf{594}, 012041 (2015).\\

\end{document}